\pdfoutput=1

\documentclass[showpacs,nofootinbib,showkeys,eqsecnum,prd,aps,preprint]{revtex4-2}

\usepackage[english]{babel}
\usepackage{comment}

\usepackage{amssymb}
\usepackage{amsmath}
\usepackage{amsfonts}
\usepackage{latexsym}
\usepackage{mathrsfs}
\usepackage{bm}
\usepackage{tensor}
\usepackage{hyperref}
\usepackage{graphicx}

\def\dac{\displaystyle\frac}

\usepackage{amsthm}

\usepackage{diagbox}

\def\dac{\displaystyle\frac}

\def\{{\lbrace}
\def\}{\rbrace}

\begin{document}

\title{Kinetics of the CuBr vapor active medium under non-typical excitation conditions}

\author{Anton E. Kulagin}
\email{aek8@tpu.ru}
\affiliation{Division for Electronic Engineering, Tomsk Polytechnic University, 30 Lenina av., 634050 Tomsk, Russia}
\affiliation{Laboratory of Quantum Electronics, V.E. Zuev Institute of Atmospheric Optics, SB RAS, 1 Academician Zuev Sq., 634055 Tomsk, Russia}

\author{Maxim V. Trigub}
\email{trigub@iao.ru}
\affiliation{Division for Electronic Engineering, Tomsk Polytechnic University, 30 Lenina av., 634050 Tomsk, Russia}
\affiliation{Laboratory of Quantum Electronics, V.E. Zuev Institute of Atmospheric Optics, SB RAS, 1 Academician Zuev Sq., 634055 Tomsk, Russia}

\begin{abstract}
We study a non-typical excitation mode of the copper bromide active medium using the kinetic model. The active medium is pumped by the pulse train and the laser generation is obtained in the subsequent single excitation pulse. Such mode allows one to obtain the laser generation pulse with the extended duration by increase in the pause duration after the pulse train. The relaxation processes during the pause are studied in order to explain such effect. It is shown that this operation mode can also be used to obtain the superradiance and amplification pulses of the extended duration that is of interest for active optical systems. Based on the comparison with the experimental results, new fundamental results are obtained regarding the copper bromide kinetics in the active medium.\\
\end{abstract}


\keywords{metal vapor active medium; kinetic model; pulse train; relaxation; copper bromide; lasing}

\maketitle

\section{Introduction}
\label{sec:int}

Improvement of frequency-energy characteristics of lasers is a relevant problem. It is associated with the overcoming technical limitations and the detailed study of physical properties of active media.Therefore, the math modelling of physical processes in active media and their experimental study are closely related to each other. The properties of active media on self-terminating transitions in metal vapors were studied in a number of both theoretical works (see, e.g., \cite{Carman199771,kushner83,marshall04,Boichenko20081522,Vuchkov1994750,Bokhan2011110,Ghani200667}) and experimental works (see, e.g., the recent works \cite{kostadinov2021,Boichenko20151,Singh2022,dikshit2022,behrouzina19} and review in \cite{batenin2017}). Moreover, there are works on implementation of non-typical excitation conditions \cite{Polunin2003833} that provide substantial change of frequency-energy and amplifying characteristics. In particular, it allows one to realize the mode with the increased duration of amplification \cite{Trigub2017828}. Such mode is useful for the visualization of remote objects with the laser monitor. In \cite{Trigub2021357}, the increased radiation pulse duration has been achieved using the pump by a train of dissociation pulses and a subsequent single excitation pulse. The stable temperature was provided by the external heating \cite{Dimaki202069} in order to maintain the stationary temperature regime when the delay between the end of the train and the excitation pulse is varied. It was shown that the laser generation is observed even when the delay between the last pulse of the train and the excitation pulse reaches 2 milliseconds. The prepulse concentrations in the active medium depend on the duration of the delay. Hence, the laser generation and electrical characteristics of the excitation pulse depends on this duration. The similar effects were studied experimentally in \cite{Trigub2017828,Gubarev201657} where the duration of the generation pulse was increased by the decrease of the repetition rate of excitation pulses. The results of \cite{Trigub2021357} are of interest from a theoretical point of view since one can judge the processes in the active medium on the scale of units of milliseconds on their basis.

This paper is devoted to the theoretical study of kinetic processes in the active medium with the extended duration of a radiation pulse that is provided by the excitation mode described above. Such study accompanied by the comparison with the experimental results allows us to obtain new data about processes that determine the operation of the copper bromide active media.

The paper is organized as follows. In Section \ref{sec:mod}, we describe our modelling approach. The operation mode of the model system is explained. In Section \ref{sec:res}, the modelling results are given and discussed. Subsection \ref{sec:res1} is devoted to the modelling of the pulse train. In Subsection \ref{sec:res2}, the relaxation processes are described based on the model results. In Subsection \ref{sec:res3}, we give the results of the modelling of the excitation pulse. The radiation and electrical characteristics of the active medium are discussed here. Subsection \ref{sec:res4} complements the results with the modelling of the superradiance mode. In Section \ref{sec:con}, we conclude with some remarks.

\section{Modelling approach}
\label{sec:mod}

The proposed approach to modelling of the copper bromide active medium is based on the kinetic model described in \cite{kulopt19}.

\begin{figure}[h]
\includegraphics[width=5.0 cm]{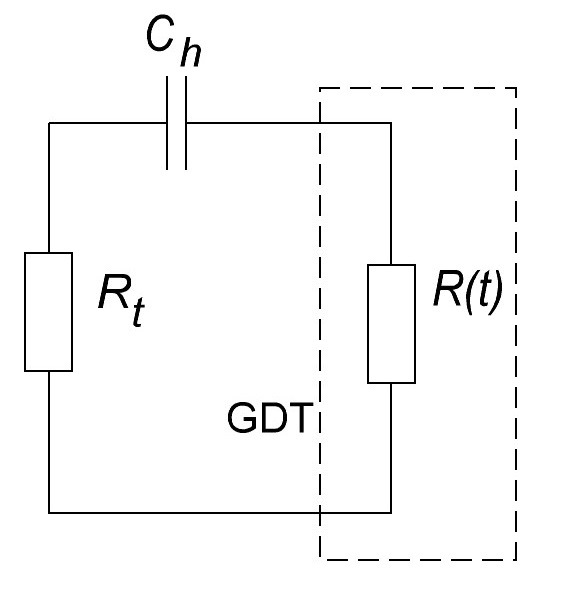}
\caption{Simplified discharge circuit for pulses of a train \label{fig1}}
\end{figure}

First, we simulate the pulse-periodic mode of operation of the active medium till the steady-state regime. It allows us to obtain the steady-state prepulse concentrations of ions and atoms in metastable states as well as the electron temperature. At this stage, the energy of pump pulses is significant but not their shape. Therefore, we use the simplest model of discharge circuit (see Fig. \ref{fig1}). The storage capacitor value $C_h$ is equal to 1100 pF. The thyratron resistance during discharge is given by
\begin{equation}
R_t=\left(1+10^6 \exp\left(-\dac{t\,{\rm [ns]}}{4}\right)\right)\, {\rm [\Omega]}.
\label{mod1}
\end{equation}
The ohmic resistance $R(t)$ of the plasma is the function of the electron temperature and concentration in the active medium. The time dependence of these plasma parameters is calculated with the kinetic model.

Note the production of copper atoms from the copper bromide is not taken into account in our model, i.e. we assume that the copper concentration is equal to its steady-state value. According to the experimental results in \cite{Astadjov19921966}, the concentration of copper and copper bromide atoms does not change at times of the order of pulse repetition period that was equal to 60 ${\rm \mu s}$ in that work. Hence, the association and dissociation of the copper bromide can be ignored in the model since these processes are much slower than other kinetic processes. Note that our model reaches the steady state after 20-30 pulse repetition periods at the frequency of 10 kHz. Thereafter, the variation of the primary prepulse characteristics of the active medium is comparable with the numerical error.

\begin{figure}[h]
\includegraphics[width=17 cm]{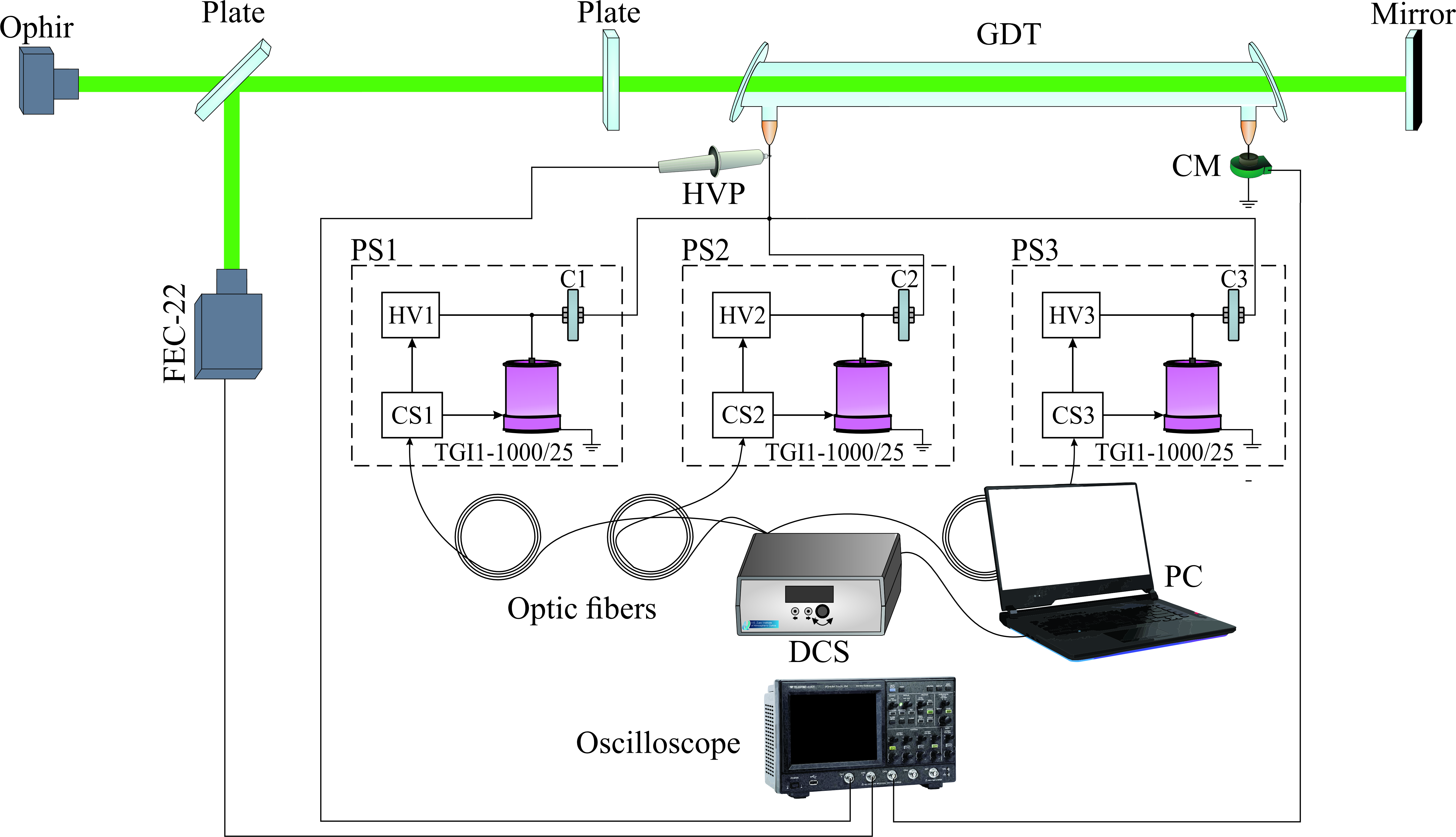}
\caption{Schematic of the experimental setup \label{figs}}
\end{figure}

After the last discharge pulse, we proceed to the modelling the relaxation of the active medium. Here, the electrical current through the gas discharge tube (GDT) is set equal to zero, and the initial conditions for the concentrations and electron temperature are assumed to be equal to the ones at the end of the last discharge pulse. The relaxation duration corresponds to the pause between the last pulse of the train and the single excitation pulse in \cite{Trigub2021357}. Some of our findings in this work will be based on the comparison of modelling the kinetics of the active medium with the experimental results obtained in \cite{Trigub2021357}. In that work, the experimental setup consists of three power supplies with pulse charging of working capacitors. The production of copper atoms was provided by a train of dissociation pulses formed by the power supply 1 (PS1). The power supply 2 (PS2) was responsible for the generation of pulses that suppress the radiation in the dissociation pulses from PS1. The amplitude of the storage capacitor voltage ($C_h$ in the model) was 14 kV in both the experiment and the model. The parameter of the pulses PS2 were selected experimentally in order to completely suppress the radiation from the PS1 \cite{Gordon1978266}. The number of pulses in the train varied from 50 to 400. The radiation pulses from PS1 were suppressed to accurately measure the radiation energy in the subsequent single excitation pulse. Hence, we model the train of the pulses from the single source instead of two pulses from PS1 and PS2 for the sake of the simplicity. The value of $C_h$ (Fig. \ref{fig1}) is selected so that mean electrical pump power was similar to the one provided by PS1 and PS2. In the experiment, after the last pulse of the train, a pause $\tau$ was made. After this pause, the active medium was pumped by the single excitation pulse formed by the power supply 3 (PS3). The pause $\tau$ corresponds to the relaxation time in the model. The schematic of the experimental setup is shown in Fig. \ref{figs}, where HV1-HV3 are high voltage sources that charge the storage capacitors and CS1-CS3 are control systems for the respective power supplies. The digital control system, which is galvanically separated by optical fiber, determines the operation mode.

During the experiment, the temperature of the GDT wall was kept constant by the external heating \cite{Dimaki202069}. Since the self-heating of the active medium is negligible under the described conditions, the gas temperature can be approximately considered constant along the GDT radius and equal to the wall temperature. In the model, it was set equal to 0.09 eV.

\begin{figure}[h]
\includegraphics[width=9.0 cm]{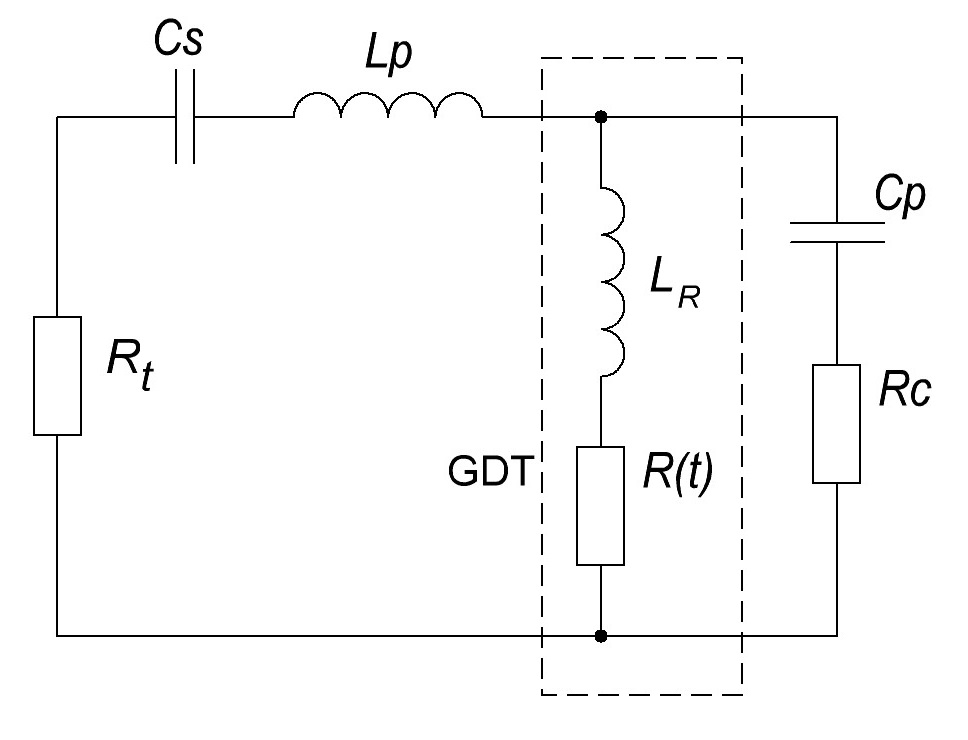}
\caption{Discharge circuit for the single excitation pulse \label{fig3}}
\end{figure}

After modelling the relaxation process with a duration $\tau$, the final values of concentrations and electron temperature in the active medium are used as new initial conditions for the excitation by a single pulse. Here, we use the more complex model of the discharge circuit that takes into account the parasitic peaking capacitance $Cp$ (see Fig. \ref{fig3}) in order to obtain the more accurate form of the pump and generation pulses. The discharge circuit in Fig. \ref{fig3} is the model of the pumping by PS3.

The capacitance of the storage capacitor $Cs=1650\,{\rm pF}$ is set equal to the experimental one in \cite{Trigub2021357} as well as its initial voltage is set equal to 14 kV. The inductance of the active medium $L_R$ corresponds to the inductance of the cylindrical conductor \cite{batenin2009} (see also \cite{Singh2022} where the similar estimate of the active medium inductance was used) and is given by
\begin{equation}
L_R=2l_d \left(\ln\left(4\dac{l_d}{d}\right)-1\right)\, {\rm [nH]},
\label{mod2}
\end{equation}
where $l_d=90$ is the length of the discharge (in cm), $d=5$ is the GDT bore diameter (in cm). The parasitic inductance $Lp$ is assumed to be equal to $L_R$. The capacitance $Cp=300\,{\rm nF}$ was evaluated based on the experimental waveforms of the current and voltage. The resistance $Rc$ is taken equal to 10 $\Omega$.

\section{Results and discussion}
\label{sec:res}

\subsection{Pulse train}
\label{sec:res1}

In this Section, we give our findings from the comparison of the modelling results with the experimental ones in \cite{Trigub2021357}.

\begin{figure}[h]
\includegraphics[width=17.0 cm]{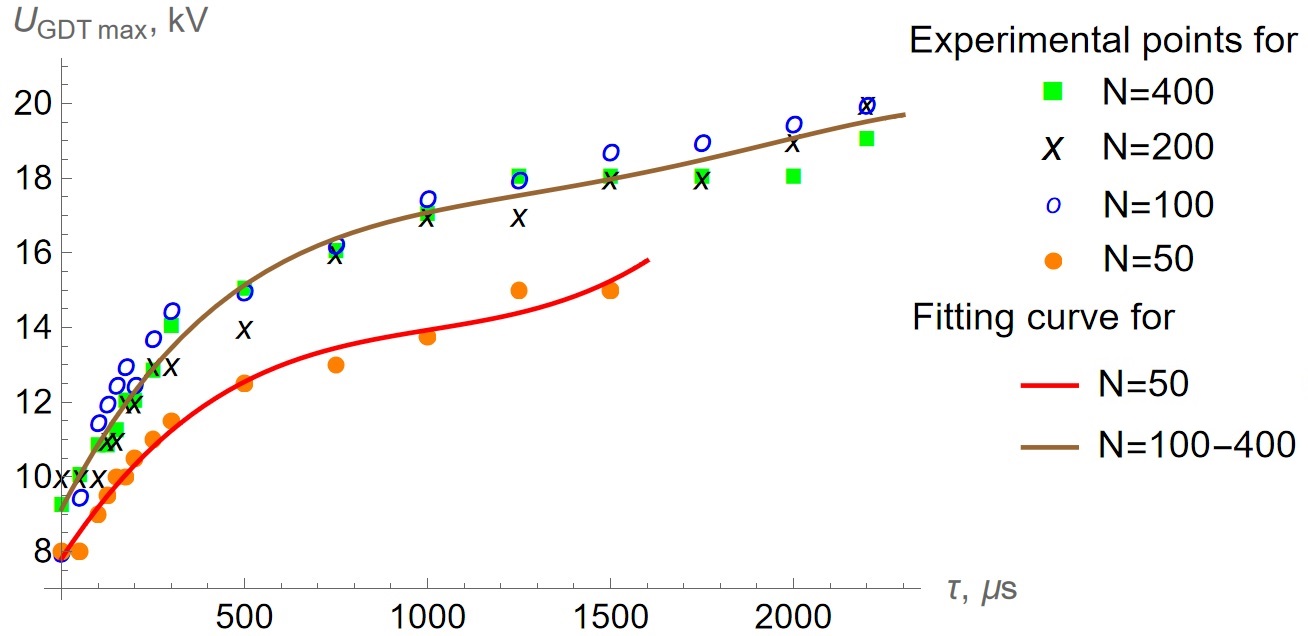}
\caption{Experimental dependence of the GDT voltage amplitude during the excitation pulse on the pulse duration for various $N$. The fitting curve for $N=100-400$ corresponds to the case when prepulse concentrations reach the steady state in a pulse train. \label{figv}}
\end{figure}

First, let us discuss the features associated with the modelling the pulse train. In our model, we assumed that active medium reaches the steady state after a pulse train. In the experiment, the difference in the GDT voltage amplitude was insignificant between modes with 100, 200, and 400 pulses in a train. Such small difference can be caused by the experimental error or the GDT operation instability. It means that the plasma resistance was almost the same in these modes. The laser generation for the mode with 100 pulses in a train differs from the modes with 200 and 400 pulses in a train for about 30\%. However, the mode with 50 pulses in a train significantly differs in both the radiation and electrical characteristics from the modes with a higher number of pulses in a train. At first glance, it contradicts our statement that prepulse concentration of ions and electron temperature reach the steady state after 20-30 pulse repetition periods. Actually, it means that there is the slow transient process which is not considered in our model. This process is the production of copper atoms from the copper bromide. Since 50 pulses are not enough to reach the steady-state concentration of copper atoms, the decrease of the average pulse repetition rate due to the decrease of the number of pulses in a train leads to the decrease of the copper atoms concentration in the excitation pulse. The experimental dependence of the GDT voltage amplitude during the excitation pulse on the pause duration $\tau$ for various number $N$ of pulses in a train is given in Fig. \ref{figv}. The polynomial fitting was used in Fig. \ref{figv}. The examples of experimental waveforms for the excitation pulse are shown in Fig. \ref{figw}.

Therefore, we can conclude that the time needed to reach the steady-state concentration of copper atoms in the copper bromide active medium exceeds 50 pulses at a pulse repetition rate of 10 kHz (i.e. exceeds 5 ms). For the mode with 100 pulses in a train, copper atoms concentration can probably considered almost steady-state since the further increase in a number of pulses in a train does not change the plasma resistance. Note that the generation energy is much more sensitive to the concentration of the active substance than the plasma resistance. Hence, even the small variation of the copper atoms concentration results in the noticeable change in the generation energy while the plasma resistance changes insignificantly.

\subsection{Relaxation stages}
\label{sec:res2}

Next, let us consider the relaxation processes in the active medium during the pause in order to explain the dependence of the radiation and electrical characteristics on the pause duration. In Fig. \ref{fig4}, the time dependencies at different distances from the GDT axis are given for the ion concentration, electron temperature, and concentration of metastable copper atoms (lower working level).


The relaxation of the active medium during the pause can be roughly divided into the following stages:\\
1) The first stage corresponds to the pause time up to about 100 ${\rm \mu s}$ when the electron gas cools down. Also, the metastable level of copper atoms decays in this stage. The recombination of ions is very slow on these time scales. Hence, the decrease in the ion and electron density is negligible in this stage. Note that the plasma of the active medium is quasineutral, i.e. the spatio-temporal distribution of the ion density and electron density are the same. Due to the decrease in the concentration of copper atoms in the metastable state, the generation energy in a single excitation pulse (PS3) increases with the increase of the pause duration within this stage.\\
2) The second stage corresponds to the pause time up to about 1000 ${\rm \mu s}$. Here, the electron gas has already cooled down and the metastable level of copper atoms has already decayed. Due to the low electron temperature, the recombination of ions proceeds rapidly. The decrease in the prepulse ion concentration for the excitation pulse leads to the increase in the plasma resistance. In this stage, there is the optimum pause duration that corresponds to the maximum generation energy in a single excitation pulse.\\
3) The third stage corresponds to the pause time of a few milliseconds. The ion recombination proceeds slowly in this stage. In the experiment, the plasma resistance significantly grew with the increase in the pause time over 1 ms. In the model, the increase in the pause time over 1 ms results in the small increase in the plasma resistance. That is the distinctive feature of this stage. In the next subsection, we will show that this feature is caused by the decrease in the concentration of copper atoms.\\

\subsection{Excitation pulse}
\label{sec:res3}

The model waveforms of the GDT voltage, current and generation obtained in a single excitation pulse after pauses of varying length are shown in Fig. \ref{fig5}.


Fig. \ref{fig5} shows that the increase in the pause duration leads to the significant increase in the duration of the generation pulse. It was also experimentally observed in \cite{Trigub2021357}. The generation pulse duration increases due to the decrease in the prepulse ion concentration. It leads to the decrease in the GDT current amplitude and the increase in the GDT current duration, i.e. the storage capacitor discharges more slowly through the GDT. In \cite{Trigub2017828}, the low energy input mode with the low GDT current of high duration was implemented at low pulse repetition rate. It also led to the extended generation pulse duration. The prerequisite for such effect is the low concentration of the atoms on the lower working level. In our case, it is realized due to the high pause duration (relaxation time), and, in \cite{Trigub2017828}, it is achieved due to the low pulse repetition rate. Note that the increase in the pause duration also changes the shape of the generation pulse. The peak of the generation pulse power moves to a latter part of the population inversion duration. The change in a shape is mainly caused by the redistribution of the energy between yellow (578.2 nm) and green (510.6 nm) radiation lines. The maximum generation energy corresponds to the pause duration of about 200 ${\rm \mu s}$ that agrees well with experimental results \cite{Trigub2021357}.

The existence of the optimum pause duration with respect to the generation energy is explained as follows. The increase in the pause duration after 200 ${\rm \mu s}$ leads to the further increase in the plasma resistance. That increases the duration of the current pulse. The increase in the current duration leads to the increase in the generation pulse duration. However, this dependence is nonlinear since there is the negative feedback in active media on self-terminating transitions that bounds the generation pulse duration \cite{Bokhan2011110}. Hence, the generation pulse duration "lags"\, behind the current pulse duration with increase in the pause duration. When the pause duration is about 200 ${\rm \mu s}$ the end of the current pulse coincides with the end of the generation pulse. When the pause duration is higher than 200 ${\rm \mu s}$ the current pulse does not end with the end of the generation pulse. The energy input after the end of the generation pulse is useless and decreases the efficient factor of the laser. When the capacitance and charge voltage of the storage capacitor are fixed, the total energy input per a pulse almost does not change. Hence, the decrease in the efficiency factor at the fixed energy input leads to the decrease in the generation pulse energy.

We have mentioned above that the conductance of the active medium almost does not change during the relaxation after 1 ms in the model. On the contrary, the amplitude of the GDT voltage increases with the increase in the pause duration over 1 ms in the experiment. When the pause duration increase from 1 ms to 2 ms, the GDT current amplitude halves in the experiment while it insignificantly changes in the model. This distinction is explained by the association of copper bromide molecules in a cold plasma. Here, the cold plasma refers to the plasma with a low electron temperature. Hence, the concentration of the active substance (copper) is starting to come down when the relaxation duration is long enough. It leads to the decrease in the plasma conductance due to the decrease in a number of atoms that can be ionized since the buffer gas (neon) atoms weakly ionize in comparison with the metal vapor ones. Based on our assumptions, we can conclude that the characteristic time of the copper bromide molecules association is a milliseconds. That also agrees with our findings above since the time needed to reach the steady-state concentration of the copper must be comparable with the characteristic time of the copper bromide molecules association. The decrease in the copper atoms concentration leads to the increase in the plasma resistance in a single excitation pulse. Also, the restoration of copper bromide molecules explains that the generation energy drops faster in the experiment \cite{Trigub2021357} with the increase in pause duration over 1 ms than in the model.

\subsection{Amplified spontaneous emission}
\label{sec:res4}

Note the high population inversion duration is of primary interest for visualization of processes from a long distance using the monostatic scheme of a active optical system when the single active medium operates as both the amplifier and the illumination source \cite{Fedorov2015}. In this case, the active medium operates without a cavity and the illumination of the visualization object is performed by the amplified spontaneous emission (ASE). Therefore, it is of interest to determine if the extended duration of the ASE (superradiance) can be obtained in such operation mode. Such possibility is not obvious since the threshold optical gain coefficient is much higher without in the absence of a cavity. For this purpose, we have additionally modelled the same pump pulse as in Fig. \ref{fig5}c but in the superradiance mode. In Fig. \ref{fig6}, the waveforms of the generation and ASE pulses are shown separately for the green and yellow radiation lines.


Although the ASE pulse is slightly shorter than one for the generation pulse, it is still much longer than the typical radiation pulse duration of 50 ns. Hence, the approach proposed is promising for the use in active optical systems. It allows one to obtain the extended duration not only for the generation pulse, but also for the ASE pulse and amplification that have been confirmed by our model results. The drawback of such approach is the impossibility to obtain a good time resolution of the visualization due to the low average pulse repetition rate.

\section{Conclusion}
\label{sec:con}

We have given the physical description of relaxation processes in the copper bromide active medium specific to the low frequency operation mode. It is shown that the relaxation can be roughly divided into three stages. The different stages are characterized by the dominance of the different transient processes. The qualitative agreement of the model results with the experiment is shown.

Based on the comparison with the experimental results, we also have shown that the association and dissociation of copper bromide molecules in the plasma of active media proceeds on time scale of a few milliseconds. In particular, in the pulse-periodic operation mode, the time needed to reach the steady state concentration of copper atoms due to the dissociation of copper bromide exceeds 5 ms, and the concentration of copper atoms begins to noticeably lower after 1 ms during the relaxation of the low temperature plasma. The lack of the direct experimental measurements of the rate of processes involving the copper bromide molecules in plasma is a big obstacle in their modelling. Therefore, the estimates obtained allow us to make a significant progress towards the mathematical description of transient processes in copper bromide active media.

Our findings indicate that the increase in the active medium resistance with the increase in duration of the preceding relaxation is caused by the decrease in the prepulse ion concentration. When the relaxation duration is higher than 1 ms, the further increase in the plasma resistance is caused by the decrease in the copper atoms concentration due the association of the copper bromide molecules.

The model results verify that it is possible to obtain the generation with the extended duration in the single excitation pulse following the train of pulses after long pause (more than 1 ms). Moreover, we showed that one can also obtain the long pulse in the superradiance mode using such pumping technique. The superradiance pulse is just about 15\% shorter than the generation pulse under the equal pump conditions. It means that such operation mode can be used to design active optical systems that require an increased duration of the superradiance and amplification.

Note that we have considered the operation mode when parameters of the discharge circuit remain unchanged with increase in the pause duration. It is of interest to study the case when energy input is varied with the pause duration due to the change of pumping conditions. Such modification will probably optimize the operation of the active medium if it can be implemented by reasonable technical means. It will be studied in our future works.
\clearpage

\section*{Acknowledgement}
We are grateful to V.A. Dimaki for the help in power supply modification, as well as N.V. Karasev and V.O. Troitskii for the discussion and interest in the work.

The study of the kinetic process in the CuBr active medium was supported by Russian Science Foundation, project no. 19-79-10096-P.
The part of the work in power supply modification was supported by Base Budget of IAO SB RAS.

\bibliography{lit1}

\clearpage

\clearpage

\begin{figure}[h]
\begin{minipage}[b][][b]{1.0\linewidth}\centering
\includegraphics[width=13 cm]{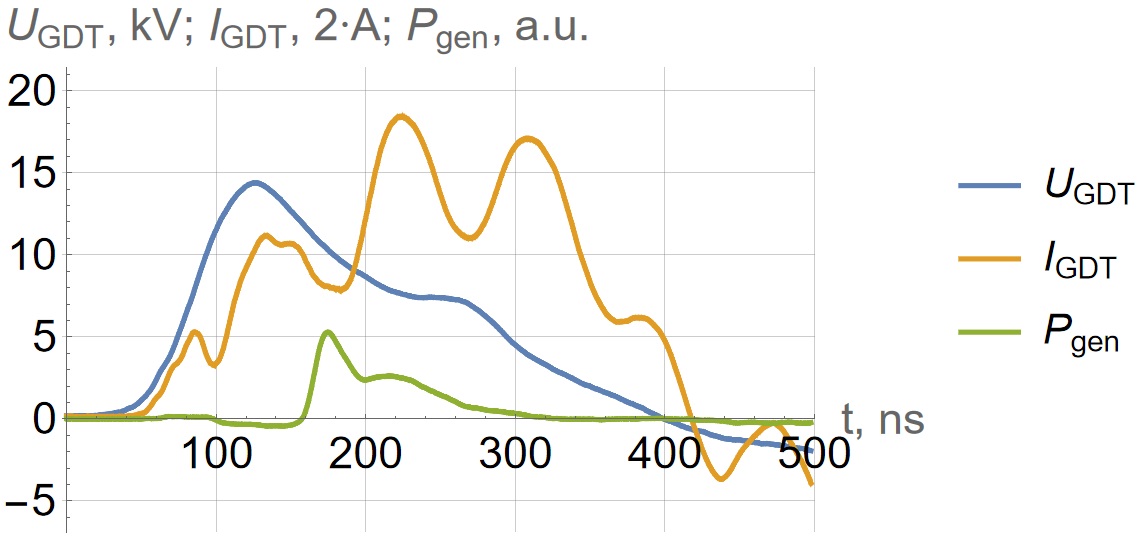} \\ a) $N=50$
\end{minipage}\\ \vfill
\end{figure}

\begin{figure}[h]
\begin{minipage}[b][][b]{1.0\linewidth}\centering
\includegraphics[width=13 cm]{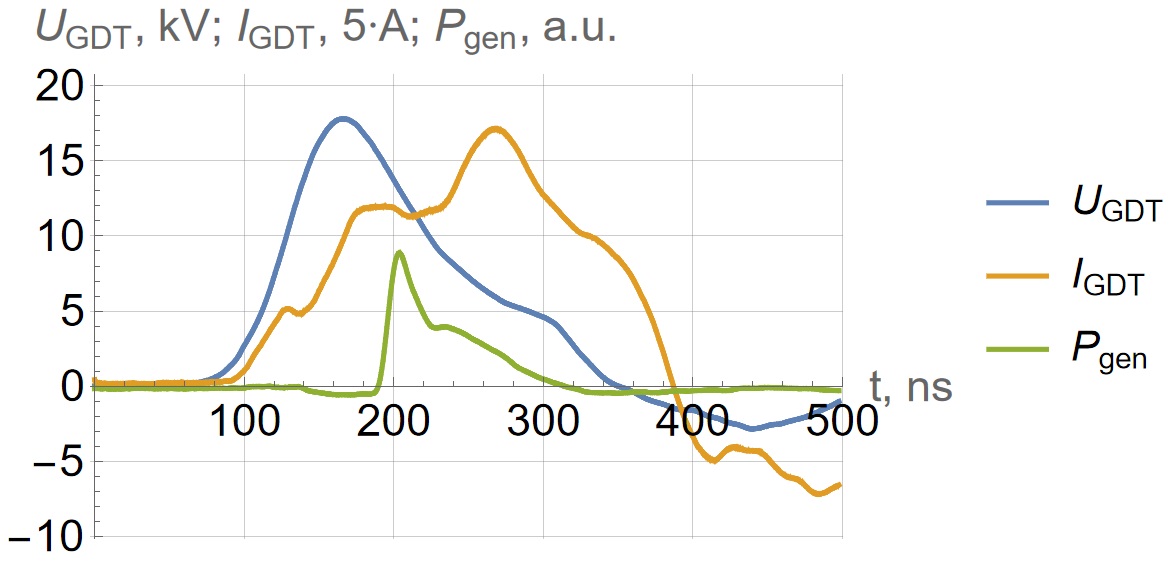} \\ b) N=200
\end{minipage}\\ \vfill
\caption{Experimental waveforms of the GDT voltage, GDT current and generation power in an excitation pulse for $N=50$ and $N=200$ for $\tau=1000\,{\rm \mu s}$. \label{figw}}
\end{figure}

\clearpage

\begin{figure}[h]
\begin{minipage}[b][][b]{1.0\linewidth}\centering
    \includegraphics[width=11 cm]{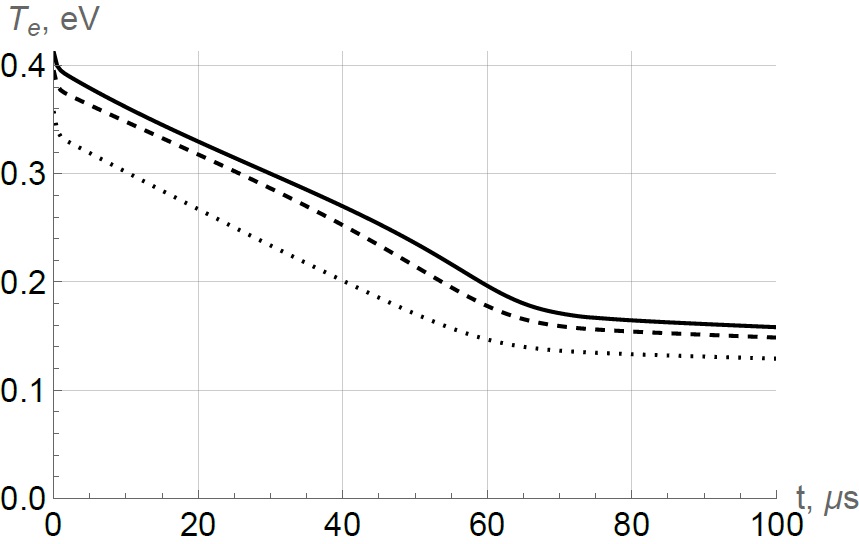} \\ a) Electron temperature
  \end{minipage}\\ \vfill
\end{figure}

\begin{figure}[h]
\begin{minipage}[b][][b]{1.0\linewidth}\centering
    \includegraphics[width=11 cm]{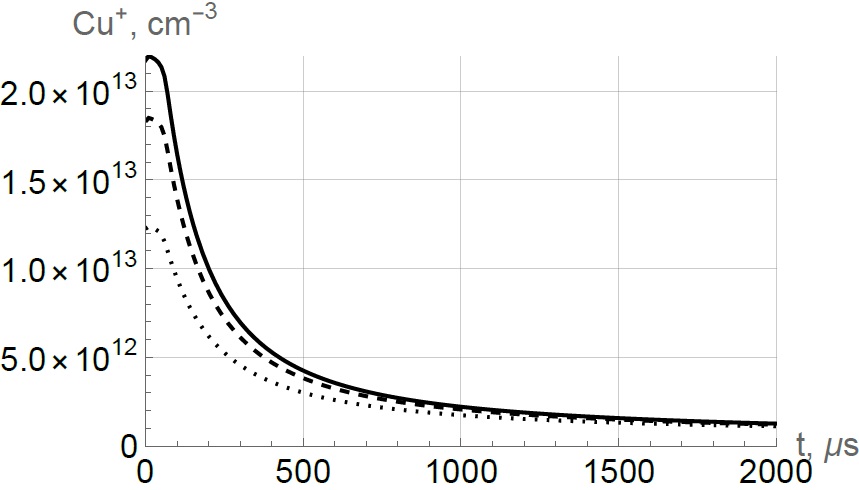} \\ b) Ion concentration
  \end{minipage}\\ \vfill
\end{figure}

\begin{figure}[h]
\begin{minipage}[b][][b]{1.0\linewidth}\centering
\includegraphics[width=11 cm]{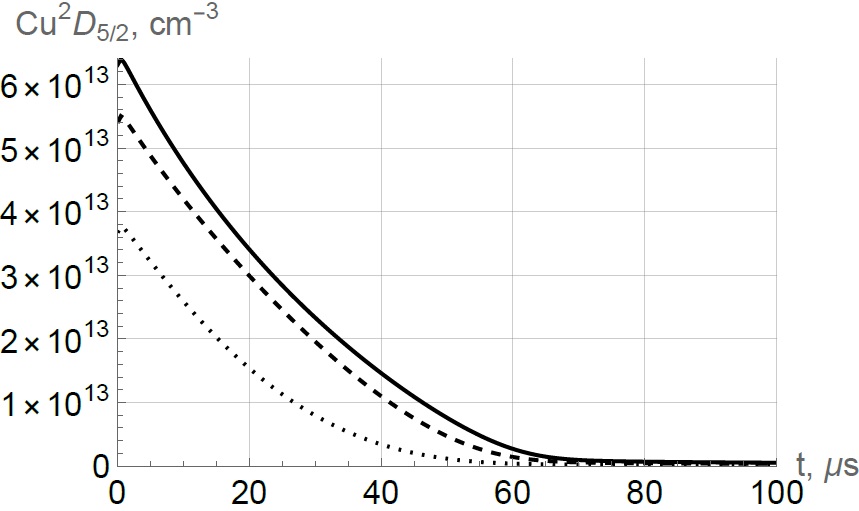} \\ c) Concentration of atoms in metastable state corresponding to the lower working level of the green radiation line (510.6 nm)
\end{minipage}\\ \vfill
\end{figure}

\begin{figure}[h]
\begin{minipage}[b][][b]{1.0\linewidth}\centering
\includegraphics[width=11 cm]{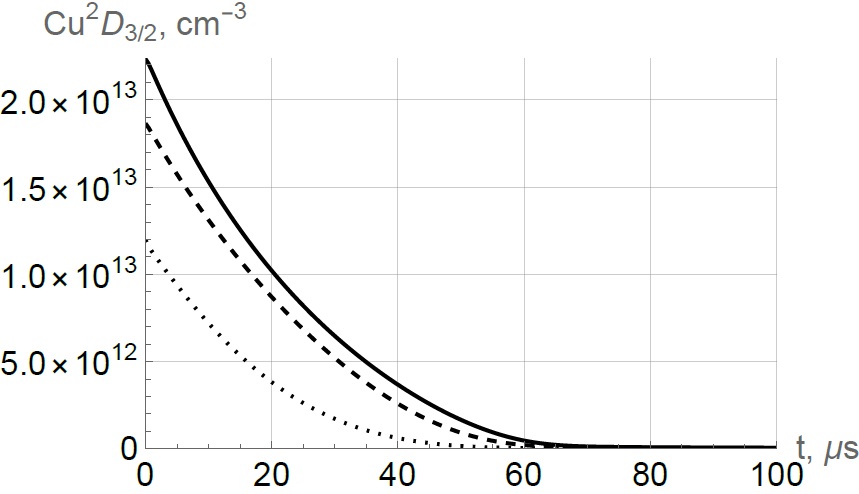} \\ d) Concentration of atoms in metastable state corresponding to the lower working level of the yellow radiation line (578.2 nm)
\end{minipage}\\ \vfill
\caption{Time dependencies of primary characteristics of the active medium during the pause. The solid line is for values on the GDT axis, the dashed line is the values at the distance of half the GDT radius, the dotted line is for the values at the distance of 80\% of the GDT radius from the GDT axis \label{fig4}}
\end{figure}

\clearpage

\begin{figure}[h]
\begin{minipage}[b][][b]{1.0\linewidth}\centering
    \includegraphics[width=12 cm]{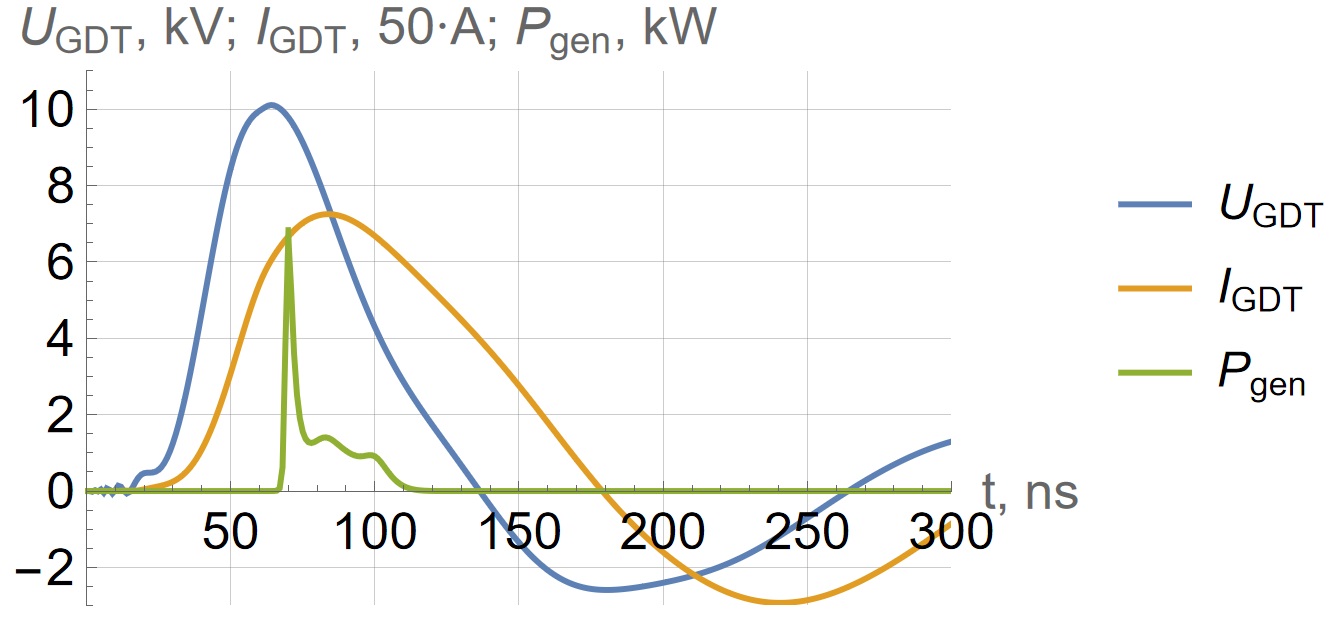} \\ a) Pause of 10 ${\rm \mu s}$
  \end{minipage}\\ \vfill
\end{figure}

\begin{figure}[h]
\begin{minipage}[b][][b]{1.0\linewidth}\centering
    \includegraphics[width=12 cm]{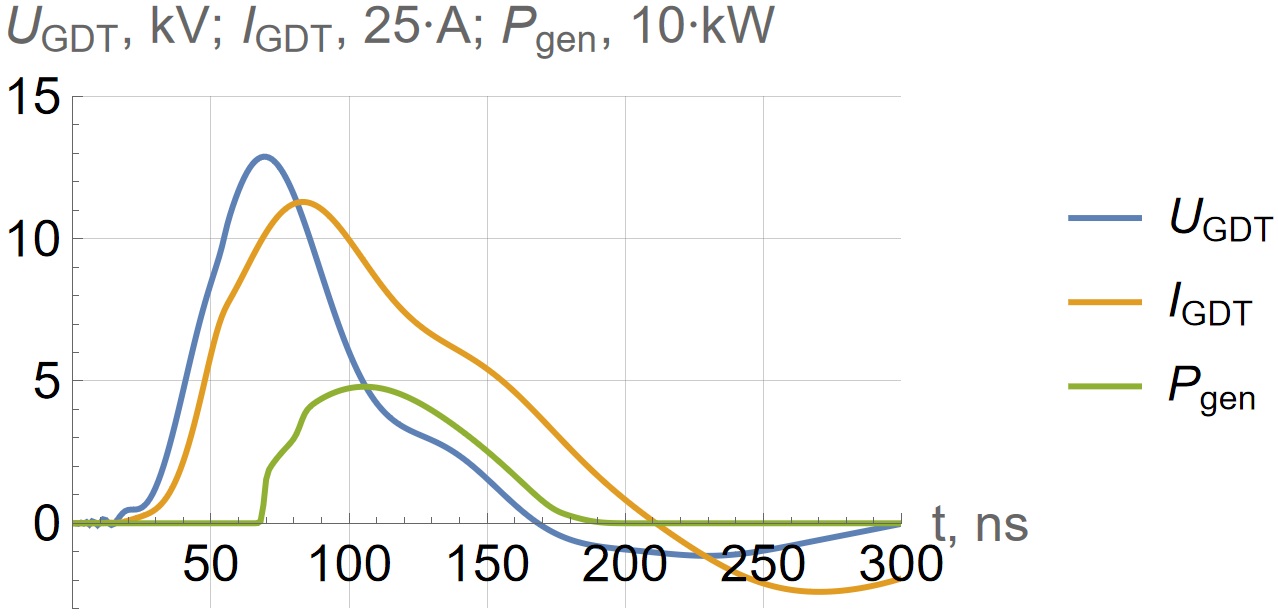} \\ b) Pause of 200 ${\rm \mu s}$
  \end{minipage}\\ \vfill
\end{figure}

\begin{figure}[h]
\begin{minipage}[b][][b]{1.0\linewidth}\centering
\includegraphics[width=12 cm]{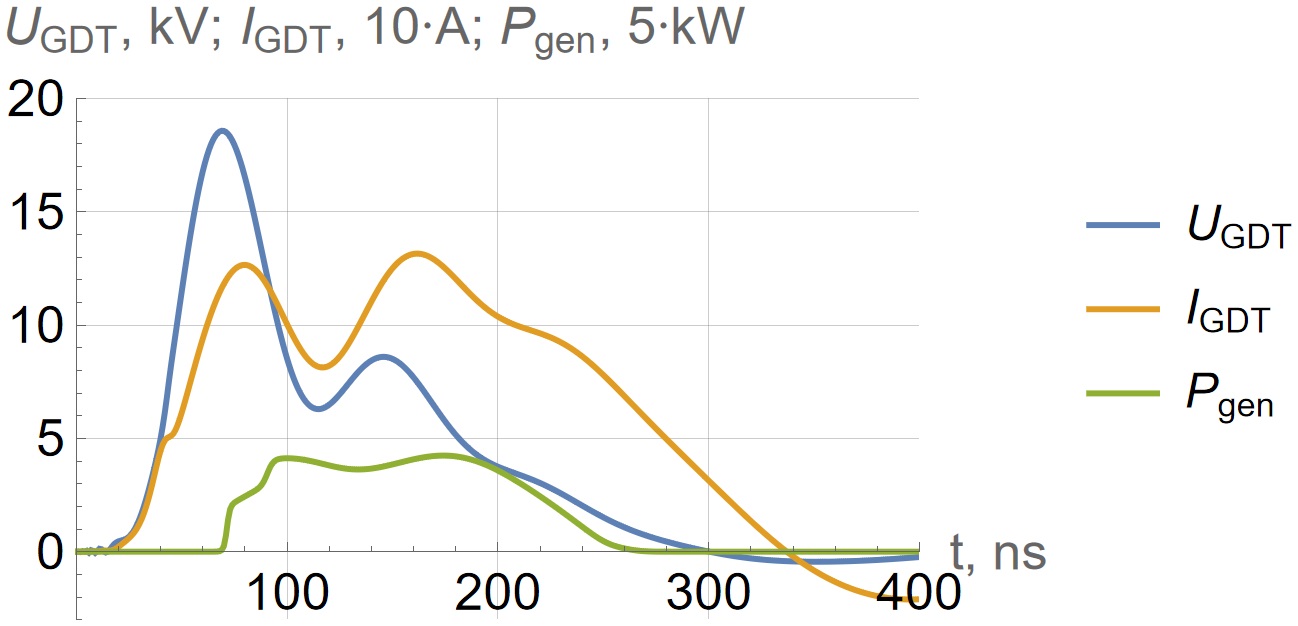} \\ c) Pause of 1000 ${\rm \mu s}$
\end{minipage}\\ \vfill
\end{figure}

\begin{figure}[h]
\begin{minipage}[b][][b]{1.0\linewidth}\centering
\includegraphics[width=12 cm]{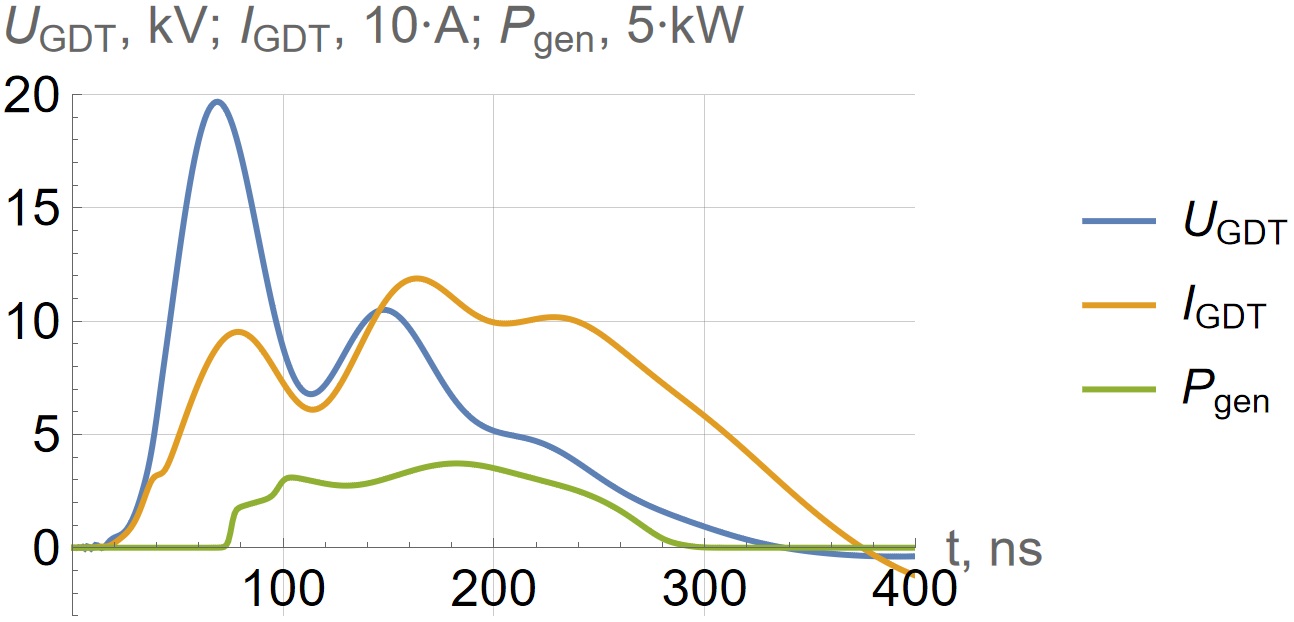} \\ d) Pause of 2000 ${\rm \mu s}$
\end{minipage}\\ \vfill
\caption{Waveforms of the GDT current, GDT voltage, and generation power for various pause durations \label{fig5}}
\end{figure}

\begin{figure}[h]
\includegraphics[width=16.0 cm]{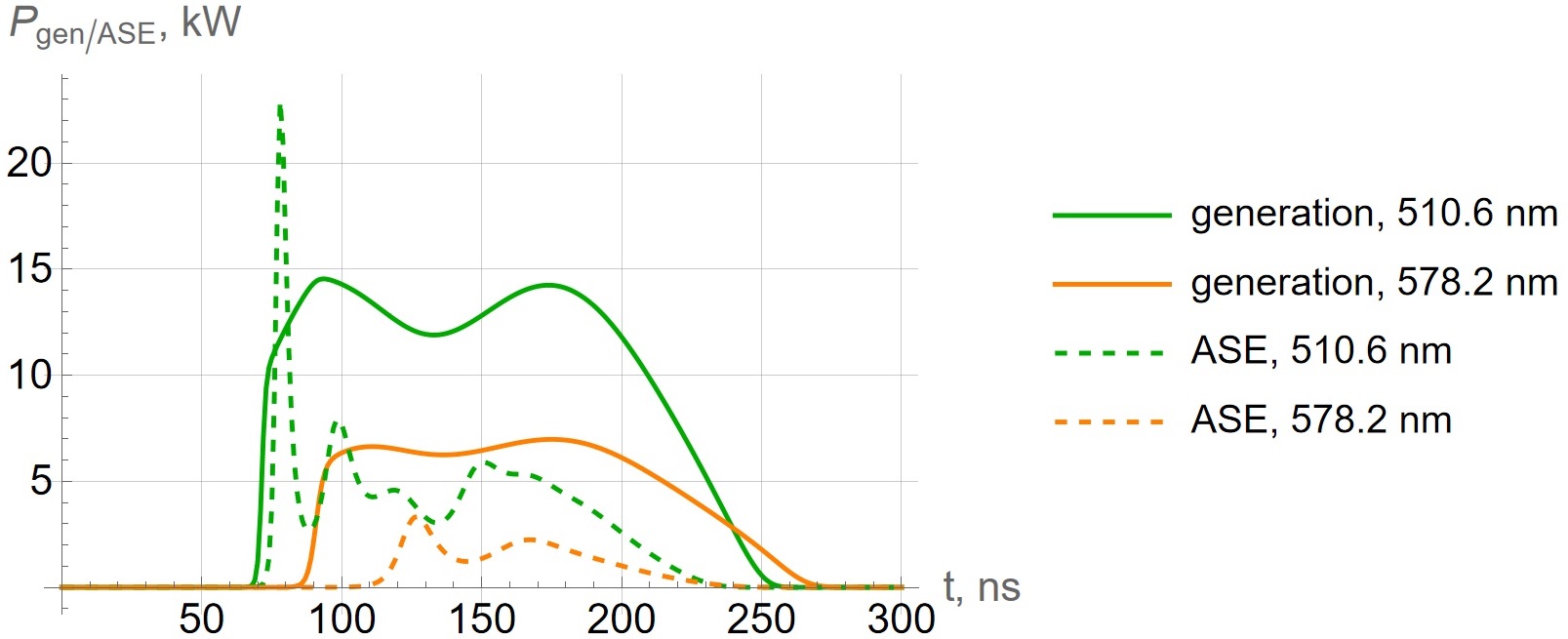}
\caption{Time dependence of the instantaneous radiation power in the single pulse after the pause of 1000 ${\rm \mu s}$. The green and yellow lines of the radiation are shown separately for the generation (with cavity) and for ASE (without cavity) \label{fig6}}
\end{figure}


\end{document}